\documentclass[%
twocolumn,
superscriptaddress,
showpacs,preprintnumbers,
amsmath,amssymb,
aps,
prb,
]{revtex4-1}

\usepackage{graphicx}% Include figure files
\usepackage{dcolumn}% Align table columns on decimal point
\usepackage{bm}% bold math
\usepackage{colortbl}
\usepackage{epstopdf}
\begin{document}
%-------------------------------------------------

%---------------------------------------------------

\title{Dynamic spin injection into a quantum well coupled to a spin-split bound state}
 %Role of the Coulomb correlations.}
% in the }
%Relaxation: quantum well-correlated impurity}
\author{N.~S.~Maslova}
\affiliation{
Lomonosov Moscow State University, 119991 Moscow, Russia}
\author{I.~V.~Rozhansky}
\affiliation{Ioffe Institute,
St.Petersburg, 194021 Russia}
\affiliation{Lappeenranta University of
Technology, FI-53851 Lappeenranta, Finland}
\author{V.~N.~Mantsevich}
\affiliation{
Lomonosov Moscow State University, 119991 Moscow, Russia}
\author{P.~I.~Arseyev}
\affiliation{
P.N. Lebedev Physical Institute RAS, 119991 Moscow, Russia}
\affiliation{
Russia National
Research University Higher School of Economics, 119991 Moscow,Russia
}
\author{N.~S.~Averkiev}
\affiliation{Ioffe Institute,
St.Petersburg, 194021 Russia}
\author{E.~L\"ahderanta}
\affiliation{Lappeenranta University of
Technology, FI-53851 Lappeenranta, Finland}
%\author{$^{1}$}
%\altaffiliation{}
%\author{$^{2}$}
%\author{$^{3}$}
%\altaffiliation{} \email{}

%\affiliation{%
%$^{1}$Ioffe Institute, 194021, St. Petersburg, Russia\\  $^{2}$
%Lomonosov Moscow State University, 119991 Moscow, Russia\\ $^{3}$
%P.N. Lebedev Physical Institute RAS, 119991 Moscow, Russia
%}%
%\affiliation{Lappeenranta University of Technology, FI-53851 Lappeenranta, Finland}

\date{\today }
\begin{abstract}
We present a theoretical analysis of dynamic spin injection
due
%by the mechanism of
to spin-dependent tunneling
%relaxation
between
%in
a quantum well (QW)
%coupled to
and a bound state
%, which is
split in spin projection due to an exchange interaction or external magnetic field. We focus on the impact of Coulomb correlations at the bound state
on spin polarization and sheet density kinetics of the charge carriers in the QW.
%Bound state and QW dynamics in the presence of Coulomb correlations
The theoretical approach
%electron kinetics is analyzed
%by means
is based on kinetic equations for the electron occupation numbers
%with the different spins and QW electrons,
taking into account high order correlation functions for the bound state electrons.
It is shown that the on-site Coulomb repulsion leads to an enhanced dynamic spin polarization of the electrons in the QW and a delay in the carriers tunneling into the bound state.
The interplay of these two effects leads to non-trivial dependence of the spin polarization degree, which can be
probed experimentally using time-resolved photoluminescence experiments.
It is demonstrated that the influence of the Coulomb interactions can be controlled by adjusting the relaxation rates.
These findings open a new way of studying the Hubbard-like electron interactions experimentally.
\end{abstract}

 \keywords{} \maketitle

\section{Introduction}
The research field of spintronics continues to grow covering various spin phenomena in solid-state physics\cite{spintronics1}.
The first generation or "metallic" spintronics is associated with magnetism and exchange interaction. It has
%this research field have
succeeded in suggesting practical applications, the vivid example is the giant magnetoresistance effect widely used in hard drives\cite{spintronicsRevModPhys}.
Semiconductor spintronics of the second generation is focused on the effects based on spin-orbit interaction, which locks a particle motion with its spin. This locking is key to many attracting practical applications\cite{Sinova2012}.
 %starting from the classical Das-Datta spin transistor.
In the traditional semiconductor materials the effective spin-orbit interaction is relatively weak, so the latest research in the field of spintronics develops in two overlapping directions: new semiconductors with stronger spin-orbit interaction, including topological insulators\cite{spint_topoNat2014,MacDonald2012}, and new spin phenomena based on the exchange interaction\cite{Parkin2010}. %, which is in many cases stronger than the spin-orbit
Our work contributes to the second direction, in the present paper we focus on a dynamical spin injection
into a quantum well (QW) through a tunnel barrier with account for the exchange interaction in the leads.

The spin injection into a semiconductor remains the cornerstone of modern spintronics.
The conductivity mismatch prevents an efficient spin injection from a ferromagnetic metal into a semiconductor\cite{Fert2001}. The widely discussed solutions of this problem (apart from those based on spin-orbit interaction) include
 %based on would be using
 using dilute magnetic semiconductor as a spin injector\cite{Weiss2014,Ohno2014}, spin injection from a ferromagnet through a tunnel barrier\cite{Rashba2000}, superdiffusive spin transport\cite{SuperDiffusive2016}.
%Other suggestions apart from those utilizing spin-orbit interaction include
%spin-filtering
%based on spin-orbit interaction\cite{our2Dholes2016},
%The spin injection
In our paper we consider a spin injection into a semiconductor due to spin-dependent relaxation of initially unpolarized ensemble of charge carriers.
In this sense it is similar to spin-depdendent recombination phenomena\cite{IvchenkoRecomb,Ivchenko2016}.
%The spin injection
%the spin-dependent relaxation of

In our model a non-equilibrium distribution of the non-polarized carriers is assumed created instantaneously in the QW. We analyse theoretically the subsequent kinetics of the spin polarization.
 The study of this physical model is motivated by the experimental studied reported in Refs.\cite{KorenevSapega,Akimov2014}. In these experiments an InGaAs based heterostrcutures with a QW and remote Mn doping layer was optically pumped with non-polarized carriers; the experimentally measured time-resolved photoluminescence indicated a development of non-stationary spin polarization of the 2D carriers in the QW on
 the time scale smaller than the radiative recombination time.
 The origin of this phenomena was explained theoretically as electron tunneling from the QW into a Mn dopant layer \cite{OUR_Spindeptun_PRB2015,ourFTP2017}. The spin splitting of the impurity bound state in the dopant layer due to exchange interaction results in the spin-dependent tunneling rate and thus leads to the spin polarization of the carriers remaining in the QW\cite{OUR_Spindeptun_PRB2015}.

     In this paper we generalize the theory by developing non-stationary formalism to describe the charge and spin kinetics in the QW coupled to a bound state.
      An important feature of our study is that in addition to the conventional Heisenberg exchange interaction
     %present
we account for the Coulomb correlations at the bound state which can contribute to the spin splitting as
it is well known from Anderson and Stoner models\cite{Hewson}.
We show how the kinetics of the spin polarization in the QW depends on the relaxation rates and the strength of the Coulomb on-site correlation at the bound state.

\section{Theoretical model}

\begin{figure}
  \centering
  \includegraphics[width=0.5\textwidth]{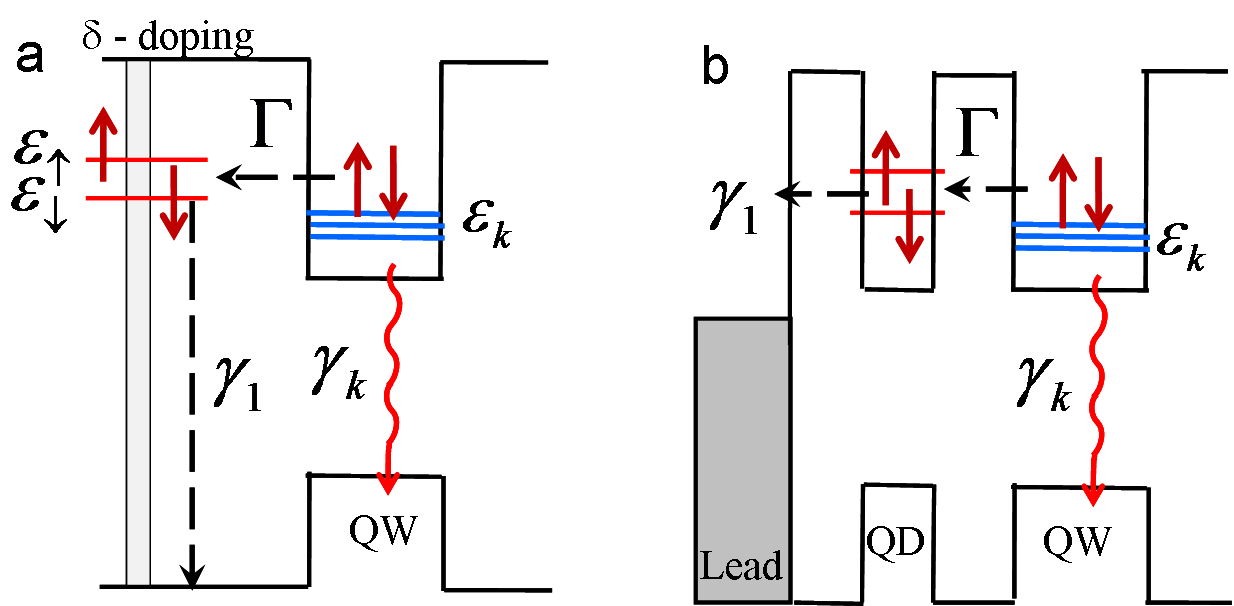}
  \caption{Two possible realizations of the considered system:
  (a) corresponds to the design of experimentally studied (Ga,In,Mn)As heterostructures with bound states formed by
  paramagnetic impurities,
  (b) is an alternative design with bound states formed by a quantum dot (QD).}
  \label{Fig1}
\end{figure}

 Although motivated by the experiments\cite{KorenevSapega,Akimov2014}, in this paper we do not restrict ourselves to a particular semiconductor heterostructure design. Let us consider two model systems shown in Fig.~\ref{Fig1}.
 Essential for both configurations is a QW with one size quantization subband. At initial time $t=0$ it is filled with unpolarized non-equilibrium 2D electrons (for example, created by an optical pumping) with the energies $\varepsilon_k$, where $k$ is the in-plane wavevector. A spin-split bound state with the energy  $\varepsilon_1$ is separated from the QW by a tunnel barrier. The barrier is characterized by the tunneling rate $\Gamma$. The radiative electron-hole recombination in the QW is characterized by the rate $\gamma_k$, which is spin-indepedent. There is also a spin-independent relaxation channel
emptying the bound state with a characteristic rate $\gamma_1$.

The model system shown in Fig.\ref{Fig1},a corresponds to the experimentally studied (Ga,In,Mn)As heterostructures\cite{KorenevSapega}, in which the bound state is formed by Mn ions in interstitial position providing donor-like states. The delta-doping Mn layer is located at a distance of 3-10 nm from the QW. The spin splitting is due to the effective exchange field in the doping Mn layer and the relaxation from the impurity donor levels is due to a very fast non-radiative recombination with the holes in the low-temperature grown (Ga,Mn)As layer\cite{OUR_Spindeptun_PRB2015,ourFTP2017}.

A somewhat different model system is shown in Fig.\ref{Fig1},b. Here the bound state is formed by a quantum dot (QD) located at a small distance from the QW. This state can accept one or two electrons.
The spin splitting inside the QD occurs due to an exchange interaction if QD has paramagnetic impurities, an external magnetic field if present and the dynamic Coulomb correlations in the QD.
The spin-independent relaxation channel is represented by a metallic lead, so that an electron can escape from QD with a characteristic rate $\gamma_1$.
The results presented below are equally valid for both schemes, Fig.~\ref{Fig1},a and Fig.~\ref{Fig1},b, moreover,
we believe that the presented theory can be applied to other systems of a similar character.
Below we make no difference between the two variants of the model system shown in Fig.~\ref{Fig1}.
%in our case is accompanied with the coloub

%Instead of In particular, we consider the system which are not to ferromagnetic

%In addition, the problem of spin injection and effective detection of spin states is %extremely important for modern spintronics.
%The idea of controlling spin polarization in semiconductor
%systems has been intensively developed since the discovery of diluted magnetic %semiconductors
%with a relatively high Curie temperature of the ferromagnetic  phase transition [1, 2]. The limitations of
%We consider non-stationary processes in the system formed by a
%quantum well coupled to a single-level impurity with Coulomb
%correlations of localized electrons. Such a system can be realized
%by the quantum well and the thin layer doped by magnetic impurities
%located a few nanometers from the quantum well.
The Hamiltonian of the system can be written in the form:
\begin{eqnarray}
\hat{H}=\hat{H}_{QW}+\hat{H}_{1}+\hat{H}_{int},
\end{eqnarray}
where $\hat{H}_{QW}$ describes the QW:
%as a sum of the quantum well part:
\begin{eqnarray}
\hat{H}_{QW}=\sum_{\sigma,k}\varepsilon_{k}c_{k\sigma}^{+}c_{k\sigma},
\end{eqnarray}
$\hat{H}_{1}$ describes the bound state with the Hubbard term for on-site Coulomb repulsion:
%the single-level impurity part
\begin{eqnarray}
\hat{H}_{1}=\sum_{\sigma}\varepsilon_{1}\hat{n}_{1}^{\sigma}+U\hat{n}_{1}^{\sigma}\hat{n}_{1}^{-\sigma},
\end{eqnarray}
and $H_T$ is the tunneling part describing the QW and bound state coupling:
\begin{eqnarray}
\hat{H}_{T}=
\sum_{k\sigma}t_{k}(\hat{c}_{k\sigma}^{+}\hat{c}_{1\sigma}+\hat{c}_{1\sigma}^{+}\hat{c}_{k\sigma}).\nonumber
\end{eqnarray}
%\textcolor{red}{QQQ}
Here index $k$ labels continuous spectrum states in the QW,
$t_{k}$ is the tunneling transfer amplitude between QW states and the bound state.
The bound state is characterized by the energy level $\varepsilon_{1}$, which
can be split due to an exchange interaction or an external magnetic field into two spin sublevels
with the energies $\varepsilon_{\sigma}=\varepsilon_1+\sigma\Delta$, where
$\sigma=\pm1/2$ is the electron spin projection and $\Delta$ is the energy gap. Operators
$\hat{c}_{k\sigma}^{+},\hat{c}_{k\sigma}$ are the
creation and annihilation operators for the electrons in the QW,
% $k$.
$\hat{n}_{1\sigma}=\hat{c}_{1\sigma}^{+}\hat{c}_{1\sigma}$ is the occupation number
operator for the bound state, operator
$\hat{c}_{1\sigma}$ destroys electron in the bound state with the spin projection $\sigma$.
%on the
%energy level $\varepsilon_1$.
$U$ is the on-site Coulomb repulsion energy
for the doubly occupied bound state.
%$impurity energy level.
Further analysis deals with the low temperature regime when the Fermi level
is well defined and the temperature is much lower than all other energy scales in the system.

\section{Non-stationary electronic transport formalism}

Let us further consider $\hbar=1$ and $e=1$ elsewhere, so the
equations of motion for the electron operators products
$\hat{n}_{1}^{\sigma}=\hat{c}_{1\sigma}^{+}\hat{c}_{1\sigma}$,
$\hat{n}_{1k}^{\sigma}=\hat{c}_{1\sigma}^{+}\hat{c}_{k\sigma}$ and
$\hat{n}_{k^{'}k}^{\sigma}=\hat{c}_{k^{'}\sigma}^{+}\hat{c}_{k\sigma}$
can be written as:

\begin{align}
i\frac{\partial \hat{n}_{1}^{\sigma}}{\partial
t}=&-\sum_{k,\sigma}t_{k}\cdot(\hat{n}_{k1}^{\sigma}-\hat{n}_{1k}^{\sigma}),
\label{1}\\
%\end{eqnarray}
%\begin{eqnarray}
i\frac{\partial \hat{n}_{1k}^{\sigma}}{\partial
t}=&-(\varepsilon_{1}^{\sigma}-\varepsilon_k)\cdot
\hat{n}_{1k}^{\sigma}-U\cdot\hat{n}_{1}^{-\sigma}
\hat{n}_{1k}^{\sigma}\nonumber\\&+
t_{k}\cdot(\hat{n}_{1}^{\sigma}-\hat{n}_{k}^{\sigma})
-\sum_{k^{'}\neq k}t_{k^{'}}\cdot\hat{n}_{k^{'}k}^{\sigma},\label{2}\\
%\end{eqnarray}
%\begin{eqnarray}
i\frac{\partial \hat{n}_{k^{'}k}^{\sigma}}{\partial
t}=&-(\varepsilon_{k^{'}}-\varepsilon_k)\cdot
\hat{n}_{k^{'}k}^{\sigma}
%\nonumber\\&-&
-t_{k^{'}}\cdot
\hat{n}_{1k}^{\sigma}
%\hat{c}_{1\sigma}^{+}\hat{c}_{k\sigma}
+t_{k}\cdot
%\hat{c}_{k^{'}\sigma}^{+}\hat{c}_{1\sigma}.
\hat{n}_{k'1}^{\sigma}.
\label{3}
\end{align}
Following the logic of Ref.~\cite{Maslova}
we substitute the solution of Eq.~(\ref{3}) into Eq.~(\ref{2})
to obtain:
\begin{align}
i\frac{{\partial \hat n_{1k}^\sigma }}{{\partial t}} =  &- (\varepsilon _1^\sigma  - {\varepsilon _k} + i\Gamma_k )\hat n_{1k}^\sigma  - Un_1^{ - \sigma }\hat n_{1k}^\sigma  + {t_k}(\hat n_1^\sigma  - \hat n_k^\sigma )\nonumber\\
&+i\sum\limits_{k' \ne k}  {t_{k'}}{t_k}\int\limits_{}^t {{e^{i({\varepsilon _{k'}} - {\varepsilon _k})\left( {t - t'} \right)}}\hat n_{k'1}^\sigma dt'} ,
\end{align}
%one can obtain the term
%$i\Gamma \hat{n}_{1k}^{\sigma}$ in Eq.~\ref{2} substituting the
%solution of Eq.\ref{3} into Eq.\ref{2}
 where
%determined as
 $\Gamma_k=\pi\nu_{0}\left(\varepsilon_k\right)t_{k}^{2}$ and $\nu_{0}\left(\varepsilon_k\right)$  is the
unperturbed density of states in the QW.
Further we assume that the tunneling parameter $t_k$ has a negligibly weak dependence on $k$, so for
2D density of states in the QW the tunneling relaxation rate is a constant
$\Gamma_k\equiv\Gamma$, which we take as a parameter.
If condition
$\frac{\varepsilon_1-\varepsilon_F}{\Gamma}\gg 1$ is fulfilled,
$\hat{n}_{1}^{\sigma}$ is a slowly varying quantity in comparison
with
%$\hat{c}_{1\sigma}^{+}\hat{c}_{k\sigma}$
$\hat{n}_{1k}^{\sigma}$. Consequently, it
is reasonable to consider that:

\begin{eqnarray}
\frac{\partial}{\partial
t}\hat{n}_{1}^{\pm\sigma}\hat{n}_{1k}^{\sigma}
%\hat{c}_{1\sigma}^{+}\hat{c}_{k\sigma}
\sim
\hat{n}_{1}^{\pm\sigma}\frac{\partial}{\partial
t}
%\hat{c}_{1\sigma}^{+}\hat{c}_{k\sigma}
\hat{n}_{1k}^{\sigma}
. \label{42}\end{eqnarray}
Taking into account that
$(\hat{n}_{1}^{\sigma})^{2}=\hat{n}_{1}^{\sigma}$ and
$(1-\hat{n}_{1}^{\sigma})\cdot \hat{n}_{1}^{\sigma}=0$ one can
find expressions for
$(1-\hat{n}_{1}^{-\sigma})\hat{n}_{1k}^{\sigma}$,
$\hat{n}_{1}^{  -\sigma}\hat{n}_{1k}^{\sigma}$ and
%, which allow one to
obtain the
equations for the time evolution of the particle number operators
$\hat{n}_{1}^{\sigma}$, $\hat{n}_{k}^{\sigma}$ for the bound state and QW, respectively.
The suggested theoretical approach allows one to analyze dynamic spin injection
due
%caused by the mechanism
%of
to the spin-dependent tunneling with
%in the  system,
%taking into
account for high order correlation functions for the bound state electrons.
Therefore, it gives possibility to analyze the effects of the on-site Coulomb repulsion.
%correctly

We also add explicitly the spin-independent relaxation terms describing recombination in the QW with the rate $\gamma_k$ and relaxation at the bound state due to non-radiative recombination (Fig.~\ref{Fig1},a) or tunnel leakage into the lead (Fig.~\ref{Fig1},b) with the rate $\gamma_1$.
Thus, we account for the effect of Coulomb correlations on the tunneling between QW and the bound state and also
%while the relaxation from the QW and the the bound state is treated
%into a lead or reservoir is treated as an effective
for the additional bound state broadening due to the relaxation into a lead or reservoir.
This approach neglects the influence of the QW and bound state relaxation channels on each other,
%Kondo-like effects,
but
allows for decoupling of the QW and bound state equations of motion.
Therefore, we obtain:
\begin{eqnarray}
\frac{\partial \hat{n}_{1}^{\sigma}}{\partial t}&=&-2\Gamma\cdot
[\hat{n}_{1}^{\sigma}-(1-\hat{n}_{1}^{-\sigma})\cdot
\hat{\Phi}\left({\varepsilon_\sigma}\right)\nonumber\\&-&\hat{n}_{1}^{-\sigma}\cdot
\hat{\Phi}\left({\varepsilon_\sigma+U}\right)]-\gamma_{1}\cdot \hat{n}_{1}^{\sigma},\nonumber\\
\frac{\partial \hat{n}_{k}^{\sigma}}{\partial t}&=&\frac{2
\Gamma}{\nu_0\pi}\cdot[(1-\hat{n}_{1}^{-\sigma})(\hat{n}_{1}^{\sigma}-\hat{n}_{k}^{\sigma})\frac{\widetilde{\Gamma}}{(\varepsilon_{\sigma}-\varepsilon_{k})^{2}+\widetilde{\Gamma}^{2}}\nonumber\\&+&
\frac{\hat{n}_{1}^{-\sigma}(\hat{n}_{1}^{\sigma}-\hat{n}_{k}^{\sigma})\widetilde{\Gamma}}{(\varepsilon_{\sigma}+U-\varepsilon_{k})^{2}+\widetilde{\Gamma}^{2}}]
-\gamma_{k}\cdot \hat{n}_{k}^{\sigma}.
\label{system0}\end{eqnarray}
%\textcolor{red}{
%$\gamma_1$ appears in $\Phi$, how it got there? There was no $\gamma_1$ in Eq 4-6
%}
Here we introduced the QW occupation operators
$\hat{\Phi}\left({\varepsilon_\sigma}\right)$ and
$\hat{\Phi}\left({\varepsilon_\sigma+U}\right)$ as:

\begin{align}
\hat{\Phi}\left({\varepsilon_\sigma}\right)&=\int
d\varepsilon_{k}\cdot \hat{n}_{k}^{\sigma}(\varepsilon_{k})\cdot\frac{1}{\pi}\frac{\widetilde{\Gamma}}{(\varepsilon_{\sigma}-\varepsilon_{k})^{2}+\widetilde{\Gamma}^{2}},\nonumber\\
\hat{\Phi}\left({\varepsilon_\sigma+U}\right)&=\int
d\varepsilon_{k}\cdot \hat{n}_{k}^{\sigma}(\varepsilon_{k}+U)\cdot\frac{1}{\pi}\frac{\widetilde{\Gamma}}{(\varepsilon_{\sigma}+U-\varepsilon_{k})^{2}+\widetilde{\Gamma}^{2}},\nonumber\\
\label{numbers0}\end{align}
where $\widetilde{\Gamma}=\Gamma+\gamma_1$.
Note, that we introduced $\widetilde\Gamma$
in order to properly account for the structure of the bound state,
which is affected both by the hybridization with the QW and the separate relaxation channel with the rate $\gamma_1$.

One can obtain equations for the bound state occupation
numbers $n_{1}^{\sigma}$  by averaging equations for the
operators and by decoupling electron occupation numbers for the
QW states from the bound state occupation numbers. Such
decoupling procedure is reasonable if one considers that Kondo
correlations can be neglected \cite{You,Zheng}. After decoupling the QW occupation numbers operators  $\hat{n}_{k}^{\sigma}$ in Eqs. (\ref{system0})-(\ref{numbers0})
have to be replaced by the distribution functions $f_{k}^{\sigma}$.
%one has to replace
%Considering
In order to take into account spin-independent relaxation processes from the QW and the bound state
%to the ground state
%as simple recombination processes one can
we add the corresponding rates $\gamma_k$ and $\gamma_1$ into kinetic
equations for the bound state occupation numbers and the QW electron
distribution function. Assuming that equilibrium
state corresponds to the empty bound state and empty QW
we obtain the following equations:
\begin{eqnarray}
\frac{\partial n_{1}^{\sigma}}{\partial
t}&=&-2\cdot\Gamma\cdot I_{k}^{\sigma}-\gamma_{1}\cdot n_{1}^{\sigma},\nonumber\\
\frac{\partial f_{k}^{\sigma}}{\partial t}&=&2\cdot \Gamma\cdot
J_{k}^{\sigma}-\gamma_{k}\cdot f_{k}^{\sigma},
\label{system}\end{eqnarray}
where

\begin{eqnarray}
I_{k}^{\sigma}&=&n_{1}^{\sigma}-(1-n_{1}^{-\sigma})\cdot
\Phi\left({\varepsilon_\sigma}\right)-n_{1}^{-\sigma}\cdot
\Phi\left({\varepsilon_\sigma+U}\right)\nonumber\\
J_{k}^{\sigma}&=&\frac{1}{\nu_0\pi}[(1-n_{1}^{-\sigma})(n_{1}^{\sigma}-f_{k}^{\sigma})\frac{\widetilde{\Gamma}}{(\varepsilon_{\sigma}-\varepsilon_{k})^{2}+\widetilde{\Gamma}^{2}}\nonumber\\&+&
\frac{n_{1}^{-\sigma}(n_{1}^{\sigma}-f_{k}^{\sigma})\widetilde{\Gamma}}{(\varepsilon_{\sigma}+U-\varepsilon_{k})^{2}+\widetilde{\Gamma}^{2}}],\label{system_1}
\end{eqnarray}
and QW occupation functions %$N_{\varepsilon}^{k\pm
%\sigma}$ and $N_{\varepsilon+U}^{k\pm \sigma}$ read
$\Phi\left({\varepsilon_\sigma}\right)$ and
$\Phi\left({\varepsilon_\sigma+U}\right)$ read

\begin{align}
\Phi\left({\varepsilon_\sigma}\right)&=\int
d\varepsilon_{k}\cdot f_{k}^{\sigma}(\varepsilon_{k})\cdot\frac{1}{\pi}\frac{\widetilde{\Gamma}}{(\varepsilon_{\sigma}-\varepsilon_{k})^{2}+\widetilde{\Gamma}^{2}},\nonumber\\
\Phi\left({\varepsilon_\sigma+U}\right)&=\int
d\varepsilon_{k}\cdot f_{k}^{\sigma}(\varepsilon_{k}+U)\cdot\frac{1}{\pi}\frac{\widetilde{\Gamma}}{(\varepsilon_{\sigma}+U-\varepsilon_{k})^{2}+\widetilde{\Gamma}^{2}}.\nonumber\\
\label{numbers}\end{align}

We further solve Eqs.~(\ref{system})-(\ref{system_1}) implying the
%with the
following initial conditions at $t=0$:
 %imfor the impurity occupation numbers
the bound state is empty, therefore
$n_{1}^{\sigma}=n_{1}^{-\sigma}=0$;
the QW is filled by the photo-excited carriers with a non-equilibrium energy distribution function characterized by chemical potential $\mu^{*}$ and electron temperature $T$:
% and describing initially
%photo-excited states
%in the QW by the model function
$f_{k}(0)=\frac{1}{e^{(\varepsilon_{k}-\mu^{*})/kT}+1}$.
%(where
%$\mu^{*}$ is a system parameter, which is not equal to the chemical
%potential) one can analyze the kinetics of impurity occupation
%numbers and QW occupation.

\begin{figure}
\includegraphics[width=80mm]{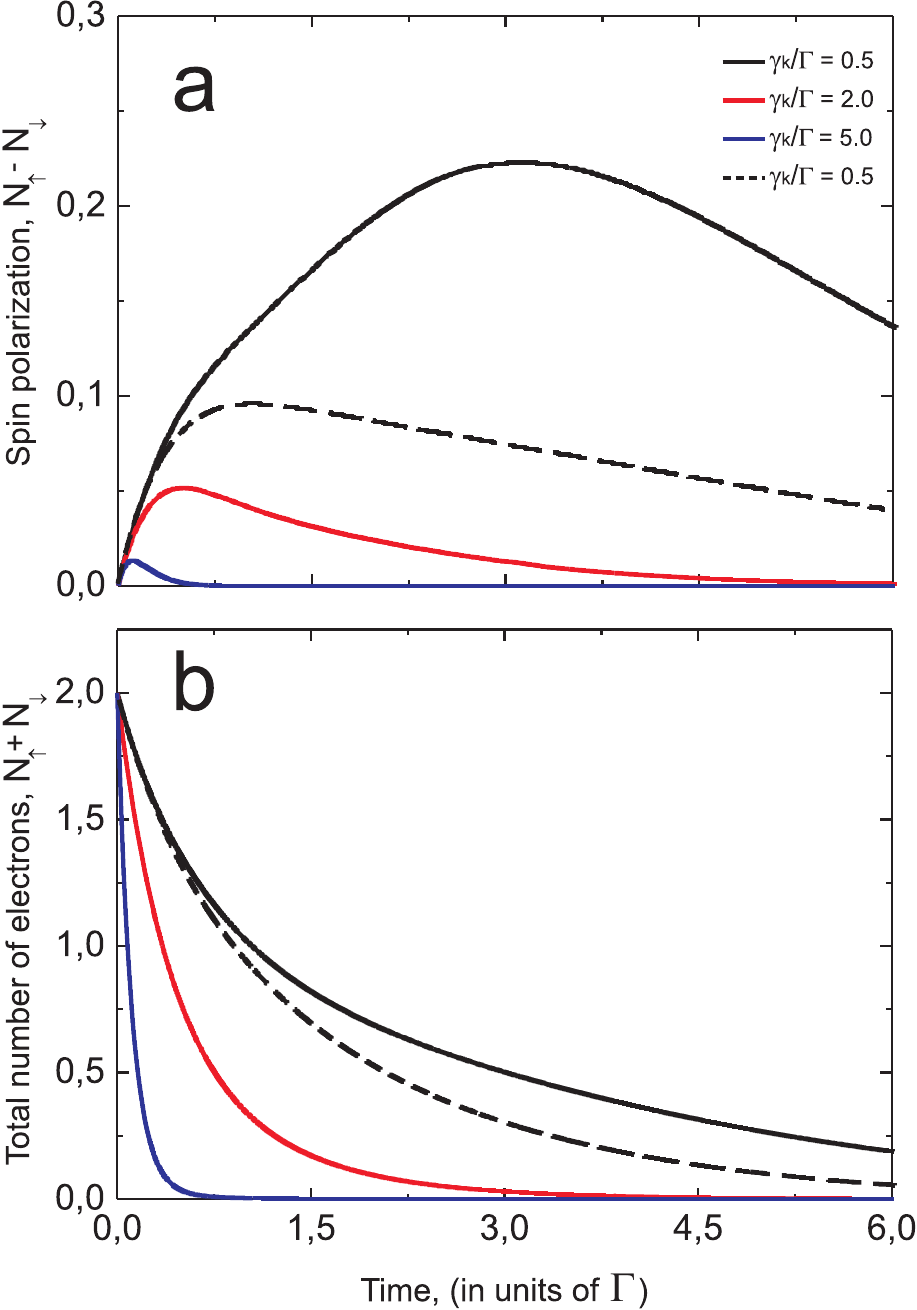}%
\caption{(Color online) Time evolution of spin polarization and QW sheet density
for different QW relaxation rates $\gamma_k$.  For
solid lines $U/\Gamma=35$, for dashed lines $U/\Gamma=0$. Parameters
$\varepsilon_\uparrow/\Gamma=2$, $\varepsilon_\downarrow/\Gamma=-2$,
$\mu^{*}/\Gamma=0$, $\gamma_1/\Gamma=0.15$ and $\Gamma=1$ are the
same for all the figures.} \label{figure2}
\end{figure}

\begin{figure}
\includegraphics[width=80mm]{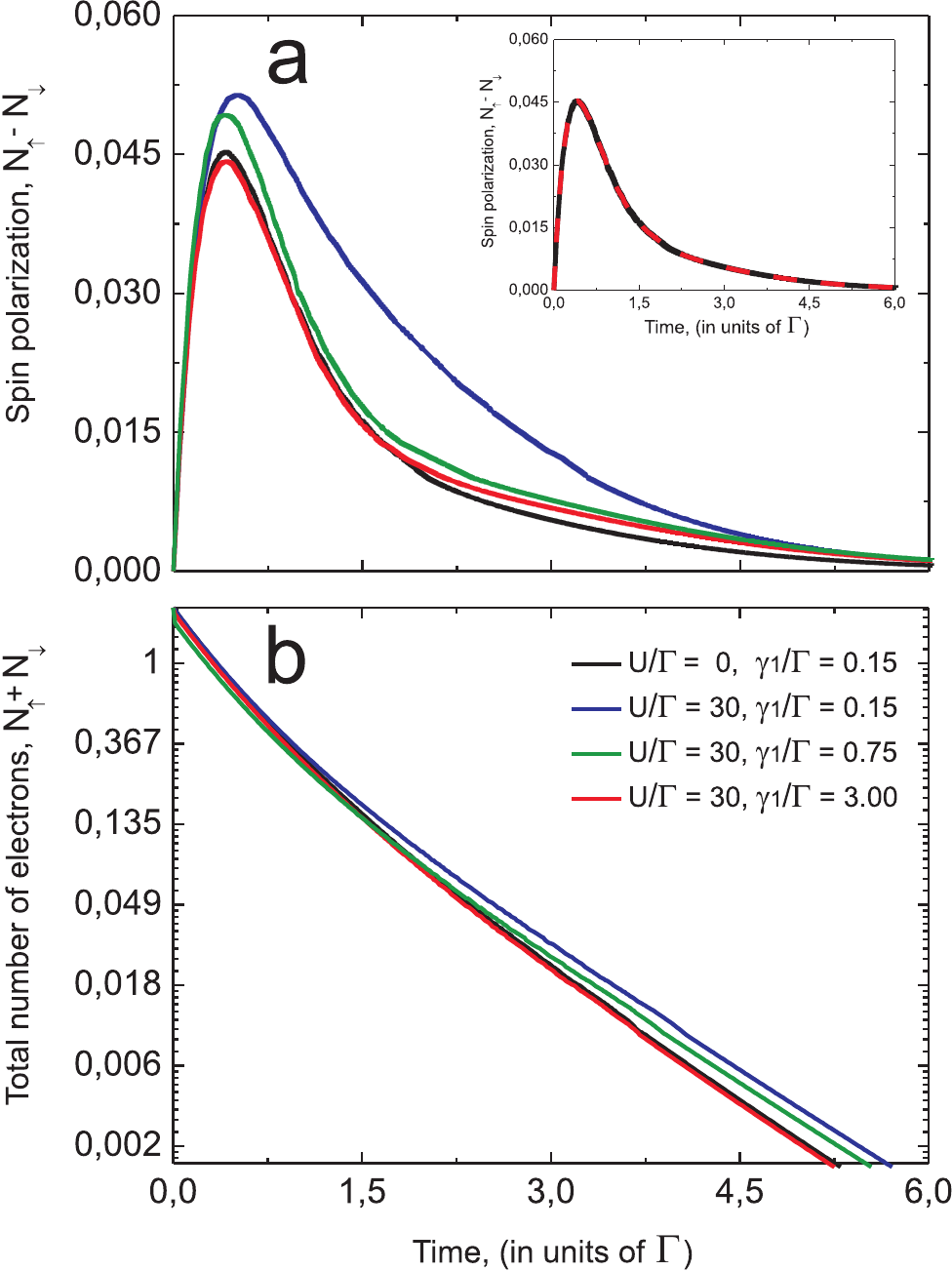}%
\caption{(Color online) Time evolution of spin polarization and QW sheet density
for different bound state relaxation rates $\gamma_1$.
Insert in
the panel (a) demonstrates that for large values of $\gamma_1/\Gamma$ spin polarization
exhibits the same behavior with (black curve) and without (red dashed curve) Coulomb interaction. 
Parameters $\varepsilon_\uparrow/\Gamma=2$, $\varepsilon_\downarrow/\Gamma=-2$,
$\mu^{*}/\Gamma=0$, $\gamma_k/\Gamma=1.5$ and $\Gamma=1$ are the
same for all the figures.} \label{figure3}
\end{figure}

\begin{figure}
\includegraphics[width=80mm]{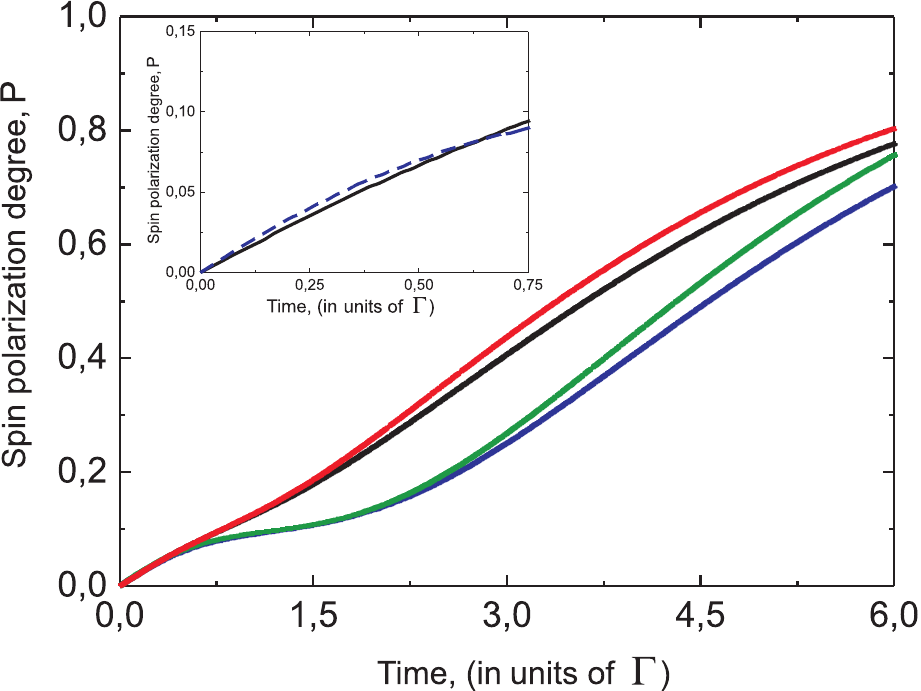}%
\caption{(Color online) Time evolution of spin polarization degree.
Parameters $\varepsilon_\uparrow/\Gamma=2$,
$\varepsilon_\downarrow/\Gamma=-2$, $\mu^{*}/\Gamma=0$,
$\gamma_k/\Gamma=1.5$ and $\Gamma=1$ are the same for all the
figures. Colors of the curves and values of parameters $U/\Gamma$
and $\gamma_1/\Gamma$ in the main panel correspond to the colors and
values shown in Fig.~\ref{figure3}. Insert shows the behavior of black
and blue curves at the beginning of the time evolution.}
\label{figure4}
\end{figure}
%Unlike the
%because it takes into account modified by the tunneling and
%relaxation processes impurity density of states.
%In the non-resonant
%case
%(for both impurity levels the following ratio takes place:
%$\frac{\varepsilon_{\sigma}-\mu^{*}}{\widetilde{\Gamma}}\gg1$ and the
%functions $I_{k}^{\sigma}$ and $J_{k}^{\sigma}$ in the right-hand part of Eqs. (\ref{system}) are of the order of
%$\frac{\widetilde{\Gamma}^{2}}{(\varepsilon_{\sigma}-\mu^{*})^{2}}$
%in comparison with the resonant case.

The spin polarization of the electrons remaining in the QW
is given by $N_{\uparrow}-N_{\downarrow}$,
where ${N_\sigma } = \int {{f_\sigma }\left( {{\varepsilon _k}} \right)d{\varepsilon _k}}$
it manifests itself
 in the circular polarization of the photoluminescence (PL) from the QW which can be measured\cite{OUR_Spindeptun_PRB2015}. %Let us
%define t
The spin polarization degree which would be measured by optical means
%in an optical experiment
is defined %of the photoluminescence circular
as:

\begin{eqnarray}
\label{eq:PolDegree}
P=\frac{N_{\uparrow}-N_{\downarrow}}{N_{\uparrow}+N_{\downarrow}}.
\end{eqnarray}
The polarization degree $P$ is negative when electrons with spin projection $\sigma=-\frac{1}{2}$ prevail.
The considered theoretical approach is more general than in Ref.\cite{OUR_Spindeptun_PRB2015}
%We would like to mention that
as Eqs.~(\ref{system}) cover both cases of
%not only
resonant and non-resonant tunneling between the QW and the bound state.

\section{Results and discussion}

The spin kinetics calculated according to the theory described above
%Calculation results are
is shown in Figs.~\ref{figure2}-\ref{figure4}.
In all the calculations we assume the same transparency
of the tunnel barrier characterized by the tunneling rate, which is taken $\Gamma=1$. Fig.~\ref{figure2} shows the time evolution of the spin polarization (a) and
total number of electrons (b) in the QW with account for the on-site
Coulomb repulsion $U$ for the electrons in the bound state. The
calculation results are presented for various relaxation rates
$\gamma_k$, which describe the radiative recombination in the QW.
%influence of the
%^Tdemonstrates that
%localized on the magnetic
%impurity
%and fixed relaxation rates $\gamma_1$ and $\Gamma$, which
%correspond to the non-radiative recombination processes in the
%magnetic layer and tunneling processes between the quantum well and
%magnetic impurity, the growth of the relaxation rate in the quantum
%well
As shown in Fig.~\ref{figure2},b, the total number of electrons in the QW is
%$\gamma_{k}$
 decreased as the initial non-equilibrium concentration of electrons relaxes
due to the tunneling and radiative recombination.
In the absence of Coulomb correlations ($U=0$), the
total relaxation rate for the electrons in the QW is simply the sum of the two $\gamma_{QW}=\gamma_k+\Gamma$.
This conclusion holds also for the case $U\ne0$ if $\gamma_k>\Gamma$. The electrons mostly relax through the recombination channel in the QW and do not have enough time to be affected by the correlations at the bound state.
This situation is illustrated by blue and red curves in Fig.~\ref{figure2}.
However, if $\Gamma>\gamma_k$ and $U\ne 0$, the
decrease of the carriers sheet density in the QW due to the tunneling is delayed as now
the tunneling of an electron requires an additional energy cost if the final state is occupied.
%Naturally, this effect becomes pronounced only when the tunneling substantially contributes to the QW total relaxation rate, i.e. $\Gamma>\gamma_k$.
This case corresponds to the solid and dashed black lines in Fig.~\ref{figure2}.

Since the bound state is split in spin projection, the relaxation rate through the tunneling channel is spin-dependent. Therefore, the spin polarization of the electrons in the QW shown in Fig.~\ref{figure2},a increases with time.
The increase is linear at $t<(\gamma_k+\Gamma)$ in agreement with Ref.~\cite{OUR_Spindeptun_PRB2015,ourFTP2017},
later on the spin polarization decays as QW becomes empty.
As can be clearly seen in Fig.~\ref{figure2},a, when the Coulomb correlations at the bound state
become important,
%However, an important finding is that when the Coulomb correlations at the bound state become important,
that is $U\ne0, \Gamma>\gamma_k$, the maximum of the spin polarization is substantially increased. For the parameters used for Fig.~\ref{figure2} the enhancement is more than two times. The position of the maximum on time scale is also substantially shifted to larger times.
Thus, the strong Coulomb correlations lead to a stronger dynamic spin injection into the QW and the delayed
kinetics, consequently, the spin polarization in the QW is preserved for a longer time.
%governed by the radiative recombination processes
%leads to the decreasing of circular polarization degree.

The effect of the Coulomb correlations on the
spin polarization in the QW
%and, consequrntly
also depends on the spin-independent relaxation rate at the bound state $\gamma_1$ (assumed to be the same for both spin sublevels).
%influence of Coulomb correlations and non-radiative recombination
This influence is shown in
%processes in the magnetic layer on the circular polarization degree
%is shown in
Fig.~\ref{figure3}.
Obviously, if the bound state sublevels are emptied faster than the rate of the incoming tunneling electrons from the QW, the Coulomb correlations shouldn't play a role as the
bound state would never get doubly occupied.  Indeed, for $\gamma_1/\Gamma>1$ the evolution of the total sheet density and the spin polarization in the QW is the same
for $U/\Gamma=30$ (red curve) and for $U=0$. In the latter case the magnitude of $\gamma_1$ does not matter as occupation of the bound state is not accompanied with an additional energy cost.
The effect of the Coulomb correlations becomes important as $\gamma_1$ is enhanced so that the electrons are less effectively removed from the bound state.
The spin injection in this case in enhanced as can be clearly seen in Fig.~\ref{figure3},a, blue line.
The QW total occupation dynamics is also affected by the Coulomb interactions, which lead to a decrease in the decay rate analogously to what was seen in Fig.~\ref{figure3},b.
However, one can note that the discrepancy between different lines in Fig.~\ref{figure3},a develops at times $t>\Gamma$. That is, when the bound state becomes significantly populated with the tunneling electrons so that the correlations become important.

%It was revealed that the presence
%of Coulomb interaction on the impurity results in the growth of
%circular polarization degree (see blue line in Fig. \ref{figure3})
%in comparison with the case when Coulomb correlations are neglected
%(see black line in Fig. \ref{figure3}). Another effect depicted in
%Fig.\ref{figure3} deals with the circular polarization degree
%decreasing with the growth of relaxation rate in the magnetic layer
%$\gamma_1$ (see green and red lines in Fig. \ref{figure3}). In the
%case when relaxation rate in the magnetic layer $\gamma_1$ exceeds
%tunneling relaxation rate $\Gamma$ and radiative recombination
%relaxation rate $\gamma_k$ the influence of Coulomb correlations on
%the kinetics of the photo-excited electrons in the quantum well and,
%consequently, on the circular polarization degree nearly vanishes.
%Large value of relaxation rate $\gamma_1$ means that correlations in
%the magnetic layer could not appear due to the fast non-radiative
%recombination processes.

Finally, Fig.~\ref{figure4} shows
the spin polarization degree $P$ introduced in Eq.~(\ref{eq:PolDegree}).
It is this quantity that can be measured experimentally as a degree of circular polarization of the photoluminescence from the QW.
Its time evolution in the presence of the Coulomb correlations is somewhat
non-trivial. The linear growth of the spin polarization degree at $0<t<\Gamma$ is common for the cases with and without on-site Coulomb correlations.
Starting from $t=\Gamma$ the increase of $P$ is suppressed by the Coulomb correlations (blue and green lines in Fig.~\ref{figure4}. This is a net effect of the two: the spin polarization, which is the nominator in (\ref{eq:PolDegree}) is enhanced due to an effectively large spin splitting of the bound state but the total occupation of QW, which is the denominator (\ref{eq:PolDegree}) remains larger as the electrons are stuck in the QW.
The Coulomb correlations do not manifest themselves if $\gamma_1\gg\Gamma$ as was discussed above (red line in Fig.~\ref{figure4}).
As the total number of non-equilibrium carriers in the system decreases the Coulomb correlations effect on polarization degree vanishes and all the curves converge at larger times in Fig.~\ref{figure4}.
%The inset in  Fig.~\ref{figure4} gives an example
%demonstrates
%that deviation from the linear growth of circular spin polarization
%degree occurs at a time $\sim1/\gamma_{k}$ due to the presence of
%Coulomb correlations (see black and blue lines in Fig.
%\ref{figure4}). Similarly to the circular polarization degree, the
%growth of relaxation rate in the magnetic layer is nearly equal to
%the absence of Coulomb correlations on the magnetic impurity (see
%black and red lines in Fig.~\ref{figure4}).
The characteristic time evolution of the polarization degree demonstrated in Fig.~\ref{figure4} has been never reported before.
In our opinion, it gives a good opportunity to verify the role of the Coulomb correlations in the systems of the considered type experimentally.
It is also clear, that the influence of the correlations can be well controlled by adjusting the system parameters.
In particular, for the system design shown in Fig.~\ref{Fig1},b the bound state relaxation rate $\gamma_1$ is directly related to the transparency of the barrier on the left, which can be tuned by changing the barrier height or its thickness.

\section{Summary}
We have studied dynamic spin injection by the mechanism of spin-dependent relaxation in a quantum well coupled to the spin-split bound state.
In this work for the first time
the impact of the Coulomb correlations
at the bound state
on the spin
%polarization and
%sheet density
and sheet density kinetics
%of the carriers
in the QW were analyzed.
As supported by our analysis, the effect of the Coulomb correlations is twofold.
Firstly, the on-site Coulomb repulsion leads to an effectively larger spin splitting and, consequently,
an enhanced spin polarization of the electrons remaining in the QW. Secondly, it increases the characteristic time of the carriers relaxation in the QW since it
%partly
%effectively
reduces the
%suppresses
%the rate of the
electron tunneling into the bound state.
We predict that the interplay of these two effects
would lead to the non-trivial dependence of a circular polarization degree of photoluminescence from the QW. This characteristic dependence will allow probing the strength of the on-site Coulomb correlations experimentally. As shown by our analysis, the effect of the Coulomb correlations can be controlled by
%adjusting the
%varying the relations between
affecting the relaxation times.
%rates.
For example, the bound state relaxation time can be tuned by a tunnel barrier separating it from the lead. This opens a way of studying the Hubbard-like electron-electron interactions experimentally.

\section{Acknowledgements}
  This  work  has  been carried out under the financial support of the
  Russian Foundation for Basic Research (RFBR) grant No. 18-02-00668
  and Scientific Program of RAS No. 9 ``Terahertz optoelectronics and spintronics''.
  V.N.M. also acknowledges the support by the RFBR grant $16-32-60024$
$mol-a-dk$.

\bibliography{manuscript_impurity}
%\begin{thebibliography}{99}

%\end{thebibliography}
 %\pagebreak

\end{document}